\documentclass[aps,prl,twocolumn,superscriptaddress]{revtex4-2}
\usepackage{amsmath,amssymb,bm}
\usepackage{graphicx}
\usepackage[colorlinks=false,pdfborder={0 0 0}]{hyperref}
\usepackage{xcolor}

\begin{document}

\title{Optical-Memory Transport Imaging: A Transport-History Framework for Finite-Memory Tracers}

\author{Haichun Liu}
\email{haichun@kth.se}
\author{Jerker Widengren}
\affiliation{Experimental Biomolecular Physics group, Department of Applied Physics,
KTH Royal Institute of Technology, S-106 91, Stockholm, Sweden}

\begin{abstract}
Finite-memory optical tracers encode upstream transport histories rather
than instantaneous local flow velocities. We introduce optical-memory
transport (OMT) imaging, a framework in which finite memory couples
internal-state relaxation to transport through memory kernels. Under
structured illumination, these histories are converted into a measurable
complex spatial-frequency response, whose local transfer-function limit
yields constant-velocity inversion. Fisher-information analysis
establishes kernel-dependent information limits and design principles
for finite-memory transport imaging. Using experimentally reported
tracer parameters, we further identify representative velocity regimes
and estimate the theoretical minimum effective photon numbers required
to reach a given velocity precision. This framework recasts finite
optical memory as a general, quantifiable resource for transport
measurement rather than a source of bias to be corrected.
\end{abstract}

\maketitle

\textbf{\textit{Introduction}}---
Optical tracers are widely used for flow imaging in fluid dynamics, soft matter physics, and biological
transport \cite{ref1}. In many tracer-based imaging approaches, measured
optical signals are implicitly treated as observables of the local
instantaneous flow state, allowing local velocity fields to be
reconstructed. Finite optical memory is ubiquitous in tracer systems
\cite{ref2}. Long-lived phosphorescence, delayed fluorescence, metastable
excited states, and multistate photophysical processes all introduce
finite temporal relaxation following optical excitation. For moving
tracers with finite optical memory, the measured optical response
depends on the excitation history accumulated along the upstream
transport trajectory prior to detection. However, the role of finite
tracer memory in determining the measured transport response has
remained comparatively unexplored. Our previous numerical investigations of photophysical structured-illumination velocimetry (PP-SIV) incorporating memory effects showed that finite tracer memory can be converted into measurable spatial phase shifts under structured illumination \cite{ref3}, but a systematic transport theory for this encoding has been lacking.

Here we introduce optical-memory transport (OMT) imaging, a general
framework in which the measured optical response is fundamentally no
longer an observable of the instantaneous local velocity field but
instead reflects the upstream transport history sampled during the
tracer memory time. Periodically structured illumination \cite{ref4}
provides a particularly direct implementation because it converts
transport-history responses into measurable complex spatial-frequency
components. In this work we report three principal advances.
\textbf{(i)} Starting from nonlinear photophysical rate equations, we
derive an effective memory kernel that couples internal-state relaxation
to advection and establish the corresponding trajectory-dependent
complex OMT response, from which the local transfer-function description
emerges as a controlled asymptotic limit. \textbf{(ii)} We formulate
velocity reconstruction as inversion of the complex OMT response and
establish the corresponding information limits for the local transfer-function approximation through
Fisher-information analysis, yielding a kernel-dependent figure of merit (FOM)
for optimizing OMT imaging experiments. \textbf{(iii)} We show that the
validity of the local transfer-function approximation is governed by a
universal dimensionless parameter, providing a quantitative validity
criterion for assessing when the full trajectory-dependent OMT forward
model may be required for quantitative flow imaging. We further
relate these theoretical results to experimentally reported
optical-tracer parameters, identifying representative velocity regimes
and theoretical lower bounds on the effective photon numbers needed for
prescribed velocity precision.

\textbf{\textit{Optical-memory transport (OMT) response theory}}---
We
consider a general tracer system, the internal photophysical state of
which is described by a vector of state populations
\(\boldsymbol{n}(t)\) under optical excitation \(I(t)\). The dynamics may in
general be nonlinear and are written as
\begin{equation}
\dot{\boldsymbol{n}}(t) = \boldsymbol{F}[\boldsymbol{n}(t),I(t)],
\label{eq:1}
\end{equation}
where \(\boldsymbol{F}\) includes arbitrary photophysical processes such as
excitation, radiative decay, nonradiative relaxation, and multistate
coupling. The measured optical signal is correspondingly given by
\(S(t) = G[\boldsymbol{n}(t)]\).

We first consider a stationary operating point under constant-intensity
excitation \(I_{0}\), for which the system reaches a steady state
\(\boldsymbol{n}_{0}\) satisfying \(\boldsymbol{F}(\boldsymbol{n}_{0},I_{0})
= 0\). A weak optical modulation is then introduced, \(I(t) = I_{0} +
\delta I(t)\), which induces a perturbation of the internal state,
\(\boldsymbol{n}(t) = \boldsymbol{n}_{0} + \delta\boldsymbol{n}(t)\).
Linearizing the nonlinear rate equations around the operating point (detailed in Supplemental Material sec. S1)
yields
\begin{equation}
\delta\dot{\boldsymbol{n}}(t) = \boldsymbol{A}\,\delta\boldsymbol{n}(t) + \boldsymbol{B}\,\delta I(t), \qquad
\delta S(t) = \boldsymbol{C}\,\delta\boldsymbol{n}(t),
\label{eq:2}
\end{equation}
where \(\boldsymbol{A}=\partial\boldsymbol{F}/\partial\boldsymbol{n}\) and
\(\boldsymbol{B}=\partial\boldsymbol{F}/\partial I\) are the Jacobian
matrices evaluated at (\(\boldsymbol{n}_{0}\), \(I_{0}\)), and
\(\boldsymbol{C} = \partial G/\partial\boldsymbol{n} \mid_{\boldsymbol{n}_{0}}\).

The resulting system is linear and time invariant (LTI) for sufficiently
small perturbations, so that standard LTI system theory applies \cite{ref5}.
The optical response therefore admits a causal convolution form,
\begin{equation}
\delta S(t) = \int_{0}^{\infty}{h(\tau)\,\delta I(t - \tau)\,d\tau},
\label{eq:3}
\end{equation}
where \(h(\tau)\) is the effective impulse-response kernel associated
with the operating point \(I_{0}\). Eq.~\eqref{eq:3} constitutes the central
reduction of the full nonlinear photophysical dynamics: the
effects of the underlying nonlinear photophysical processes are
incorporated into the operating-point-dependent response kernel, while
the measured modulation response remains linear in the small-signal
limit. The explicit form of the memory kernel follows directly from the
solution of the linear system, Eq.~\eqref{eq:2}:
\begin{equation}
h(\tau) =
\begin{cases}
\boldsymbol{C}e^{\boldsymbol{A}\tau}\boldsymbol{B}, & \tau \geq 0, \\
0, & \tau < 0,
\end{cases}
\label{eq:4}
\end{equation}
where \(e^{\boldsymbol{A}\tau}\) denotes the
matrix exponential of the Jacobian \(\boldsymbol{A}\), and the lower
branch follows from causality. Stability of the
photophysical dynamics requires all eigenvalues of \(\boldsymbol{A}\) to
have negative real parts.

We now consider moving tracers advected through a spatially varying
optical field \(I(\boldsymbol{r})\). For a tracer detected at position
\(\boldsymbol{r}\) and time \(t\), the optical excitation experienced at an
earlier time \(t - \tau\) is determined by the upstream trajectory of
the tracer, \(\boldsymbol{r}_{-\tau} \equiv \boldsymbol{r}(t - \tau)\).
The temporal response described by Eq.~\eqref{eq:3} therefore becomes a
trajectory-dependent transport response,
\begin{equation}
\delta S(\boldsymbol{r},t) = \int_{0}^{\infty}{h(\tau)\,\delta I(\boldsymbol{r}_{- \tau})\,d\tau},
\label{eq:5}
\end{equation}
so that the optical response at position \(\boldsymbol{r}\) is
determined by the excitation history sampled along the upstream
trajectory \(\boldsymbol{r}_{- \tau}\), rather than by the instantaneous
local flow state.

For steady deterministic flow fields, the upstream trajectory is uniquely
determined by the velocity field through
\begin{equation}
\frac{d\boldsymbol{r}_{- \tau}}{d\tau} = - \boldsymbol{v}(\boldsymbol{r}_{- \tau}), \qquad \boldsymbol{r}_{0} = \boldsymbol{r}.
\label{eq:6}
\end{equation}
The measured response therefore depends on the finite upstream history
sampled during the memory time rather than solely on the instantaneous
local flow velocity. The details of the complete derivation of the OMT
framework are given in the Supplemental Material sec. S1.

\textbf{\textit{Complex OMT response under structured illumination}}---
Periodically structured illumination provides a direct spatial
frequency-domain implementation of OMT imaging. Without loss of
generality, we consider a one-dimensional excitation grating \(I(x) =
I_{0} + I_{1}\cos(qx)\), where \(q\) is the spatial modulation wavevector
and \(I_{1} \ll I_{0}\) ensures that the small-signal response derived
above remains valid.

Substituting this excitation into the trajectory-dependent OMT response,
Eq.~\eqref{eq:5}, the response can be written exactly in complex form as
\begin{equation}
\delta S(x) = I_{1}\,\mathrm{Re}\left[e^{iqx}H_{\mathrm{eff}}(x)\right],
\label{eq:7}
\end{equation}
where
\begin{equation}
H_{\mathrm{eff}}(x) = \int_{0}^{\infty}{h(\tau)\exp\{iq[x_{- \tau}(x) - x]\}\,d\tau}.
\label{eq:8}
\end{equation}
Eq.~\eqref{eq:7} is the fundamental readout equation of
structured-illumination OMT imaging: it shows that while the emission remains
spatially localized at the detection position \(x\), its complex
modulation coefficient is determined by the upstream transport history
sampled along \(x_{-\tau}(x)\) and encoded through the optical-memory
kernel. Eq.~\eqref{eq:8} therefore defines
the trajectory-dependent complex OMT response, which serves as the
central forward quantity throughout the OMT imaging framework.

Since the trajectory-dependent OMT response \(H_{\mathrm{eff}}(x)\) is complex,
Eq.~\eqref{eq:7} admits the equivalent polar representation,
\begin{equation}
\begin{aligned}
&\delta S(x) = I_{1}A_{\mathrm{eff}}(x)\cos[qx - \phi_{\mathrm{eff}}(x)], \\
&A_{\mathrm{eff}} = |H_{\mathrm{eff}}|,\ \ \phi_{\mathrm{eff}} = -\arg H_{\mathrm{eff}}.
\end{aligned}
\label{eq:9}
\end{equation}
The complete derivation of the complex OMT response
under structured illumination is detailed in Supplemental Material
sec. S2.

\begin{figure*}
\includegraphics[width=0.85\textwidth]{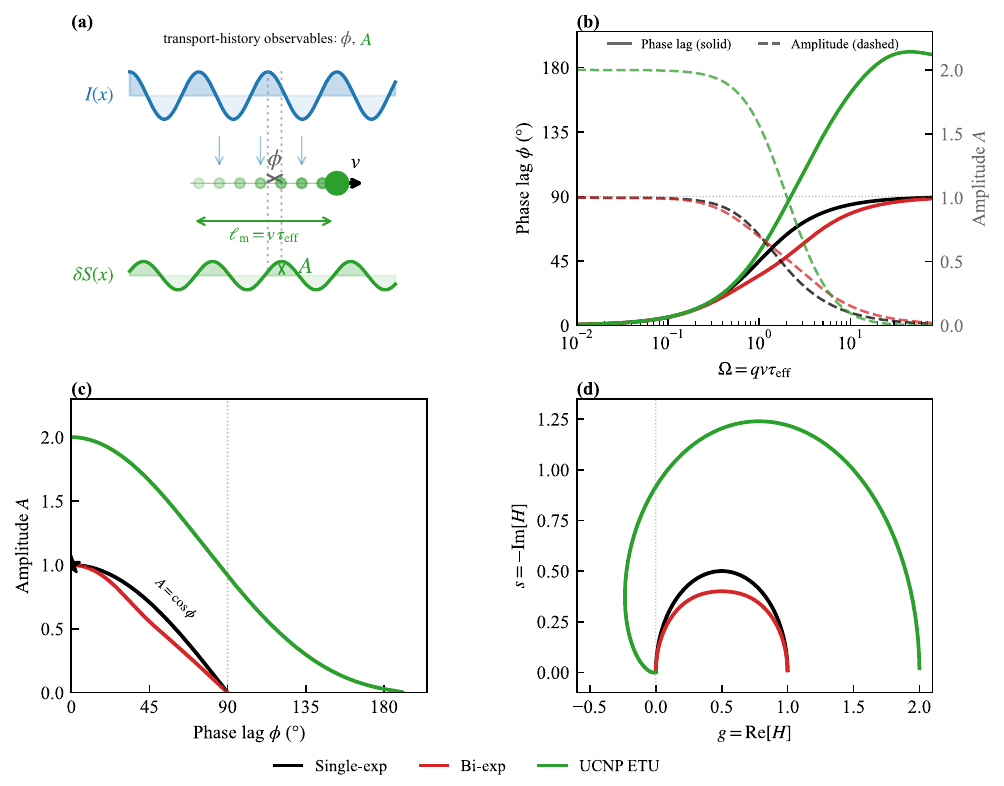}
\caption{\textbf{OMT framework and complex response of optical-memory tracers under structured illumination.} \textbf{(a)} Concept of the OMT framework. \textbf{(b)} Phase lag \(\phi(\mathrm{\Omega})\) and modulation amplitude \(A(\mathrm{\Omega})\) as functions of the dimensionless transport-memory parameter \(\mathrm{\Omega} = qv\tau_{\mathrm{eff}}\), for three representative optical-memory kernels: single-exponential (single-exp) (black), bi-exponential (bi-exp) (red, with \(w_{1} = w_{2} = 0.5\), \(\tau_{2}/\tau_{1} = 4\), Supplemental Material
	sec. S4), and UCNP ETU (green, with the dimensionless operating-point parameter \(r\) =1, Supplemental Material
	sec. S5) kernels, where \(q\) is the illumination wavevector, \(v\) the tracer velocity, and \(\tau_{\mathrm{eff}}\) the effective memory time of the optical-memory kernel. The UCNP ETU kernel is not restricted to the 0-90° phase range of positive exponential kernels and may exhibit phase excursions exceeding 180°, owing to the negative contribution in its transfer function decomposition. \textbf{(c)} Amplitude--phase loci in the \(A - \phi\) plane for the three representative kernels, where the single-exponential kernel satisfies \(A = \mathrm{cos(\phi)}\). \textbf{(d)} Representative raw OMT phasor trajectories for the three representative kernels, where the single-exponential kernels trace the universal semicircle \((g-\tfrac{1}{2})^{2}+s^{2}=\tfrac{1}{4}\).}
\label{fig:1}
\end{figure*}

Under structured illumination, transport-history responses are converted
into experimentally measurable complex response coefficients. The response magnitude characterizes the
emission modulation amplitude, while its phase describes the spatial shift relative to the excitation grating, illustrated in
Fig.~\ref{fig:1}a.

\textbf{\textit{Local transfer functions of OMT imaging}}---
The
trajectory-dependent complex OMT response derived above applies to
arbitrary steady deterministic transport. A particularly important limit is
locally uniform transport, in which the tracer experiences an
approximately constant velocity over the optical-memory time, so that
the upstream trajectory is locally approximated by \(x_{-\tau}(x) = x -
v\tau\). Substituting this into the trajectory-dependent complex OMT
response, Eq.~\eqref{eq:8}, gives
\begin{equation}
\begin{aligned}
&H_{\mathrm{eff}}(x) = \int_{0}^{\infty}{h(\tau)}e^{- iqv\tau}\,d\tau \equiv H(qv), \\
&H(\omega) \equiv \int_{0}^{\infty}{h(\tau)}e^{- i\omega\tau}\,d\tau,
\end{aligned}
\label{eq:10}
\end{equation}
where \(H(\omega)\) is the ordinary one-sided transfer function of the
optical-memory kernel, following the standard definition for
continuous-time LTI systems \cite{ref5} (Supplemental Material, sec. S3).
Locally uniform transport therefore converts the trajectory-dependent
OMT response into the ordinary transfer function of the optical-memory
kernel, evaluated at the transport frequency \(\omega = qv\) to which
tracer motion is mapped under structured illumination.

For a single-exponential kernel \(h(\tau) =
\tau_{\mathrm{m}}^{-1}e^{-\tau/\tau_{\mathrm{m}}}\), where \(\tau_{\mathrm{m}}\) is the memory time
(for emitters exhibiting single-exponential relaxation, the memory time
corresponds to the conventional fluorescence or luminescence lifetime),
the transfer function and its polar components are
\begin{equation}
\begin{aligned}
&H(\omega) = \frac{1}{1 + i\omega\tau_{\mathrm{m}}}, \qquad
A(\Omega) = \frac{1}{\sqrt{1 + \Omega^{2}}}, \\
&\phi(\Omega) = \tan^{-1}\Omega, \qquad \Omega \equiv qv\tau_{\mathrm{m}},
\end{aligned}
\label{eq:11}
\end{equation}
with \(\Omega\) the dimensionless transport-memory parameter. As \(\Omega\) increases, the modulation amplitude decreases
monotonically from unity to zero, whereas the phase increases
monotonically from 0 to 90\textdegree{} (Fig.~\ref{fig:1}b).

The single-exponential kernel provides the fundamental reference case.
More generally, finite-memory optical tracers may exhibit multiple
relaxation processes, leading to more complex optical-memory kernels.
For an arbitrary kernel normalized to unit
zero-frequency gain, \(H(0)=1\), it is useful to introduce the
effective memory time and the corresponding transport-memory parameter (Supplemental Material, sec. S4),
\begin{equation}
\tau_{\mathrm{eff}} = \int_{0}^{\infty}\tau h(\tau)\,d\tau, \qquad
\Omega \equiv qv\tau_{\mathrm{eff}},
\label{eq:12}
\end{equation}
which provide a common dimensionless transport scale for comparing
different kernel families. For responses with non-unit
zero-frequency gain, such as the raw ETU response retained in
Fig.~\ref{fig:1}, \(\tau_{\mathrm{eff}}\) can instead be defined from
the low-frequency phase delay,
\(\tau_{\mathrm{eff}} = -d\arg H/d\omega|_{\omega=0}\). For kernels with
\(H(0)=1\), this definition reduces to Eq.~\eqref{eq:12} (Supplemental
Material, sec. S5). For kernels lacking a single characteristic
relaxation time, the transfer function is evaluated as a function of the
transport frequency \(\omega = qv\), while the results are presented
versus the transport-memory parameter \(\Omega\).

We then compare the local transfer functions for three representative
optical-memory kernels: a single-exponential kernel, a bi-exponential
kernel, and a two-photon energy-transfer upconversion (ETU) kernel
representative of lanthanide upconversion nanoparticles (UCNPs) \cite{ref6}.
For the bi-exponential kernel (Supplemental Material
sec. S4), the modulation amplitude remains bounded
between 0 and 1, while the phase remains within the range
0--90\textdegree{} (Fig.~\ref{fig:1}b), similar to the single-exponential
case. Different relative amplitudes of the two exponential components
nevertheless generate distinct transfer-function curves. By contrast,
the ETU kernel (Fig. S1, Supplemental Material,
sec. S5) exhibits qualitatively different behavior,
where both the modulation amplitude and the phase are no longer
constrained by the limits of positive exponential kernels (Fig.~\ref{fig:1}b),
owing to its nonlinear energy-transfer dynamics.

The polar representation \((A,\phi)\) provides an
interesting complementary description of the same complex transfer
function. For the single-exponential kernel, eliminating the
transport-memory parameter \(\Omega\) yields the universal trajectory
\(A = \cos\phi\) in the amplitude-phase plane (Fig.~\ref{fig:1}c). For
the bi-exponential kernel, different combinations of the
involved two relaxation times and their weights generate distinct
trajectories in the amplitude-phase plane, but all lying inside the
corresponding single-exponential envelope (Fig.~\ref{fig:1}c). In contrast, the
ETU amplitude-phase trajectory extends beyond the amplitude and phase
manifold accessible to positive exponential kernels (Fig.~\ref{fig:1}c), owing to
the non-convex structure of the ETU transfer function (Supplemental
Material sec. S5).

The same responses may equivalently be visualized in the phasor plane
(\(g = \mathrm{Re}(H)\), \(s = -\mathrm{Im}(H)\)), analogous to fluorescence lifetime phasor analysis \cite{ref7}, but with the position determined by the
transport-memory parameter rather than fluorescence lifetime. Representative phasor trajectories for different kernel families are shown in Fig.~\ref{fig:1}d. The single-exponential response traces the universal semicircle
\((g-\tfrac{1}{2})^{2}+s^{2}=\tfrac{1}{4}\) and positive bi-exponential
kernels remain inside it, whereas the non-convex ETU response can extend
outside this region. Their geometric interpretation is discussed in more detail in
Supplemental Material sec. S6.

\textbf{\textit{Velocity inversion in OMT imaging}}---
The preceding
sections establish the forward model of OMT imaging. Under structured
illumination, tracer transport histories are encoded into the
trajectory-dependent complex OMT response. In the locally uniform
transport limit, this response reduces to the local transfer function of
the optical-memory kernel. Velocity imaging therefore becomes an inverse
problem in which the experimentally measured complex OMT response,
denoted by \(H_{\mathrm{obs}},\) is compared with the corresponding theoretical
forward model. Experimentally, the complex response can be recovered from
the excitation and emission patterns using several equivalent
procedures, such as peak-displacement method, spatial phase fitting, quadrature fitting, or Fourier demodulation
(Supplemental Material sec. S7).

For the single-exponential optical-memory kernel, the local transfer
function is analytically invertible, using either the phase or the
modulation amplitude (Eq.~\eqref{eq:11}) for \(v \geq 0\):
\begin{equation}
v = \frac{\tan\phi}{q\tau_{\mathrm{m}}} = \frac{1}{q\tau_{\mathrm{m}}}\sqrt{\frac{1}{A^{2}} - 1}.
\label{eq:13}
\end{equation}
As both observables originate from the same complex transfer function,
they provide equivalent velocity estimates for nonnegative
velocities. For signed velocities, the phase additionally encodes the
flow direction, whereas the amplitude depends only on \(|v|\).

For general optical-memory kernels, analytical inversion is generally
unavailable because the velocity cannot be
expressed explicitly in terms of analytical phase or amplitude relations.
Nevertheless, as the forward
model remains completely specified through the complex transfer
function, velocity reconstruction can therefore be implemented via numerical
inversion. For a prescribed kernel model, the measured complex
response \(H_{\mathrm{obs}}\) can be fitted to the corresponding theoretical transfer
function, simultaneously utilizing both modulation amplitude and phase
(Supplemental Material sec. S7).

The local transfer-function inversion developed above applies only under
locally uniform transport. For spatially varying flows, velocity
reconstruction is performed by fitting the experimentally measured
complex response directly to the trajectory-dependent OMT response described in Eq.~\eqref{eq:8}
under a prescribed flow parameterization (Supplemental Material sec. S7).

\textbf{\textit{Information limits and optimal operating
conditions}}---
The existence of an analytical or numerical velocity
inversion does not imply uniform reconstruction accuracy. Even when the
velocity is uniquely recoverable, the available information depends
strongly on the operating condition and on the sensitivity of the
memory-kernel transfer function to velocity. We quantify this sensitivity using the
Fisher information and the corresponding Cramér--Rao bound \cite{ref8}.

We model the demodulated complex response as \(H_{\mathrm{obs}} = H(v) +
n\), where \(n\) is zero-mean complex Gaussian noise with equal
quadrature variance \(\sigma^{2}\) (Supplemental Material sec. S8). Writing
\(H = Ae^{-i\phi}\), the Fisher information for the scalar local velocity
is
\begin{equation}
\begin{aligned}
F(v) &= \frac{1}{\sigma^{2}}\left|\frac{\partial H}{\partial v}\right|^{2} \\
&= \frac{1}{\sigma^{2}}\left[\left(\frac{\partial A}{\partial v}\right)^{2} + A^{2}\left(\frac{\partial\phi}{\partial v}\right)^{2}\right]
\equiv F_A(v) + F_\phi(v).
\end{aligned}
\label{eq:14}
\end{equation}
Eq.~\eqref{eq:14} retains both amplitude and phase information and
constitutes the Fisher-information
expression when the absolute response scale is known. In practice,
uncertainty in the effective spatial modulation transfer or its
calibration can introduce an unknown positive real amplitude factor into
the measured complex response, without changing its phase. Treating this
factor as a nuisance parameter removes the amplitude contribution to the
effective Fisher information exactly under the adopted noise model,
leaving (Supplemental Material, sec. S8):
\begin{equation}
F_{\mathrm{eff}}(v) = F_{\phi}(v) = \frac{A^{2}}{\sigma^{2}}\left( \frac{\partial\phi}{\partial v} \right)^{2}.
\label{eq:15}
\end{equation}
Equation~\eqref{eq:14}
applies when the absolute response scale is independently calibrated.
With the
photon-shot-noise normalization adopted (Supplemental Material sec.
S8),
\begin{equation}
\sigma^{2} = \frac{1}{N}, \qquad F_{\mathrm{eff}}(v) = NA^{2}\left(\frac{\partial\phi}{\partial v}\right)^{2},
\label{eq:16}
\end{equation}
where \(N\) is the effective photon number of the
demodulated local response, defined through the quadrature noise
variance \(\sigma^{2}=1/N\); for a background-free, uniformly sampled
sinusoidal fringe of modulation depth \(m\) under shot-noise-limited detection, \(N \simeq m^{2}N_{\mathrm{det}}/2\), where
\(N_{\mathrm{det}}\) is the number of detected photons in the analysis
region (see Supplemental Material, sec. S8). From the Cramér--Rao inequality \(\mathrm{Var}(\hat{v}) \geq 1/F_{\mathrm{eff}}(v)\), where \(\hat{v}\) denotes the velocity estimator,
we obtain that
\begin{equation}
\frac{\sigma_{v}}{v} \geq \frac{1}{A(\Omega)\left|v\,\partial\phi/\partial v\right|\sqrt{N}},
\label{eq:17}
\end{equation}
where \(\sigma_{v}\equiv\sqrt{\mathrm{Var}(\hat v)}\), giving the
photon-limited relative-precision bound for velocity estimation. Using \(\Omega =
qv\tau_{\mathrm{eff}}\) (with \(q\) and \(\tau_{\mathrm{eff}}\) fixed, so
that \(v\,\partial\phi/\partial v = \Omega\, d\phi/d\Omega\)), Eq.~\eqref{eq:17}
becomes
\begin{equation}
	\frac{\sigma_{v}}{v} \geq \frac{1}{A(\Omega)\left|\Omega\, d\phi/d\Omega\right|\sqrt{N}}.
	\label{eq:18}
\end{equation}
For general kernels, the phase derivative is evaluated numerically with
respect to the transport frequency \(\omega = qv\) (Supplemental
Material sec. S8).

\begin{figure*}
	\includegraphics[width=0.75\textwidth]{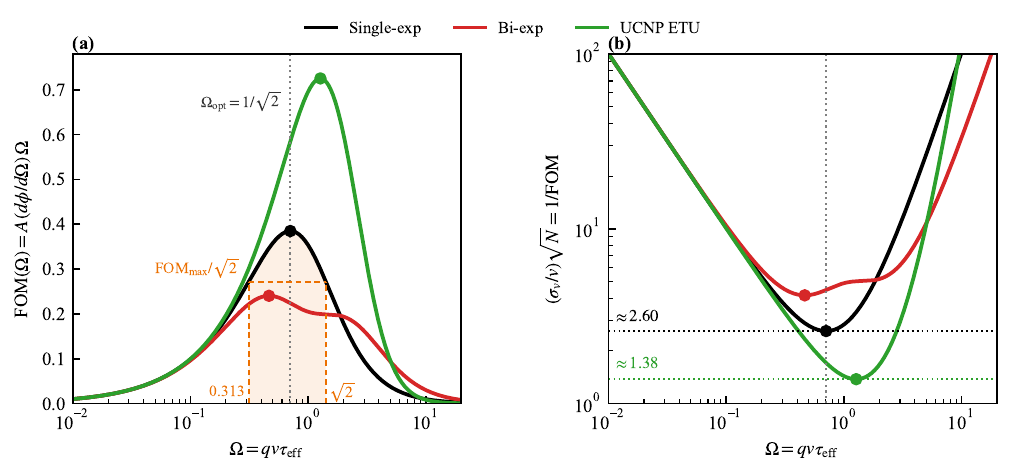}
	\caption{\textbf{Fisher-information analysis of OMT velocity estimation under the local transfer-function approximation.} \textbf{(a)} Dimensionless velocity-sensing figure of merit, \(\mathrm{FOM}(\Omega) = A(\Omega)\Omega\left. \mid d\phi/d\Omega \right.\mid\), obtained from the optical-memory transfer function and plotted as a function of the transport-memory parameter \(\Omega = qv\tau_{\mathrm{eff}}\), for three representative optical-memory kernels (single-exponential, bi-exponential, and optimized UCNP ETU). The vertical line indicates the single-exponential optimum \(\Omega_{\mathrm{opt}}^{\mathrm{SE}} = 1/\sqrt{2}\). \textbf{(b)} Relative velocity precision floor (Cramér--Rao bound, CRB), \(\sqrt{N}\,\sigma_{v}/v = 1/{\mathrm{FOM}(\Omega)}\), as a function of \(\Omega\). The plotted minima are approximately 2.60 for the single-exponential kernel and 1.38 for the optimized UCNP ETU kernel in the weak-upconversion limit. The CRB is calculated from the corresponding transfer-function analysis and presented versus the transport-memory parameter. The orange-shaded region in (a) marks the high-sensitivity range of the single-exponential kernel, \(\mathrm{FOM} \geq \mathrm{FOM}_{\max}/\sqrt{2}\) (\(0.313 \lesssim \Omega \leq \sqrt{2}\)), within which the photon number required for a given CRB is at most twice its minimum [Eq.~\eqref{eq:19}].}
	\label{fig:2}
\end{figure*}

Eq.~\eqref{eq:18} naturally identifies the dimensionless velocity-sensing
FOM and the corresponding photon-shot-noise-limited
relative precision (Supplemental Material sec. S8),
\begin{equation}
\mathrm{FOM}(\Omega) = A(\Omega)\,\Omega\left|\frac{d\phi}{d\Omega}\right|, \qquad
\frac{\sigma_{v}}{v} \geq \frac{1}{\mathrm{FOM}(\Omega)\sqrt{N}}.
\label{eq:19}
\end{equation}
For the single-exponential kernel, \(\mathrm{FOM}_{\mathrm{SE}}(\Omega) =
\Omega/(1+\Omega^{2})^{3/2}\), which vanishes at both small and large
\(\Omega\): at small \(\Omega\), the phase is weakly sensitive to
velocity, whereas at large \(\Omega\), the modulation amplitude is
strongly attenuated. Their balance produces the optimum
\begin{equation}
\begin{aligned}
&\Omega_{\mathrm{opt}}^{\mathrm{SE}} = \frac{1}{\sqrt{2}}, \qquad
\mathrm{FOM}_{\max}^{\mathrm{SE}} = \frac{2}{3\sqrt{3}} \simeq 0.385, \\
&\left.\frac{\sigma_{v}}{v}\right|_{\min}^{\mathrm{SE}} \simeq \frac{2.60}{\sqrt{N}},
\end{aligned}
\label{eq:20}
\end{equation}
(Fig.~\ref{fig:2}a and Supplemental Material sec. S8).

Different kernel families generate distinct FOM landscapes (Fig.~\ref{fig:2}a).
For kernels without closed-form expressions, the FOM is obtained
numerically from the corresponding transfer functions (Supplemental
Material, sec. S8) and presented using the normalized transport
frequency \(\Omega\) (Eq.~\eqref{eq:12}). For the representative
bi-exponential kernel, the second relaxation process broadens and shifts
the response but lowers the maximum FOM relative to the
single-exponential reference. By contrast, numerical optimization of the
UCNP ETU kernel gives \(\mathrm{FOM}_{\max}^{\mathrm{ETU}} \simeq 0.725\)
when the sensitizer and emitter decay rates are approximately matched
and in the
weak-upconversion limit (Supplemental Material sec.
S8). Varying the operating point
\(r\), related to excitation intensity,
 changes the FOM only weakly, from
\(\mathrm{FOM}_{\max} \simeq 0.556\) at \(r=0.01\) (near the strong-upconversion limit) to \(\simeq 0.575\) at
\(r=0.99\) (near the weak-upconversion limit) (Supplemental Material sec. S8). The principal improvement therefore arises from the
rate-ratio-dependent ETU kernel shape rather than from the
operating-point parameter \(r\) alone.

The kernel-dependent FOMs translate directly into different
velocity-precision floors (Fig.~\ref{fig:2}b): at its optimal
operating point, the optimized ETU kernel gives \(\left.\sigma_v/v\right|_{\min}^{\mathrm{ETU}} \simeq 1.38/\sqrt{N}\),
corresponding to a 1.88-fold improvement over the single-exponential
floor of Eq.~\eqref{eq:20} at fixed effective photon number \(N\). This result provides a concrete kernel-engineering target:
matching the sensitizer and emitter decay rates enhances velocity
information without increasing the photon budget.

The kernel-dependent optimum also provides a direct
structured-illumination design rule. For a target velocity \(v\) and
effective memory time \(\tau_{\mathrm{eff}}\),
\begin{equation}
\Lambda_{\mathrm{opt}} = \frac{2\pi v\tau_{\mathrm{eff}}}{\Omega_{\mathrm{opt}}},
\label{eq:21}
\end{equation}
which reduces to \(\Lambda_{\mathrm{opt}}^{\mathrm{SE}} = 2\pi\sqrt{2}\,v\tau_{\mathrm{m}}\)
for the single-exponential kernel.

\textbf{\textit{Validity of the local transfer-function approximation}}---

The preceding analysis established both the forward transfer function
and the corresponding information limits under the assumption of locally
uniform transport. We now determine the conditions under which this
local transfer-function approximation remains valid and when the full
trajectory-dependent OMT response may instead be required.

Eq.~\eqref{eq:8} shows that the optical response detected at position \(x\)
is determined by the excitation sampled along the upstream trajectory
\(x_{-\tau}(x)\). Because the response kernel \(h(\tau)\) decays over a characteristic memory time, only a finite segment of this trajectory contributes
appreciably to the measured signal. This sampled trajectory segment
constitutes the spatial manifestation of tracer optical memory.

Using the effective memory time introduced in Eq.~\eqref{eq:12}, we therefore define
the characteristic memory length as the distance travelled by the tracer
over the effective support of the response kernel, \(l_{\mathrm{m}} \sim
|v(x)|\,\tau_{\mathrm{eff}}\), and compare it with the distance over
which the velocity field varies appreciably along the upstream
trajectory, quantified by the local streamline-variation scale
\(L_{v}(x) = |v(x)| / |dv/dx|_{x}\).
The analysis in Supplemental Material sec. S9 predicts that the validity
of the local transfer-function approximation is governed by the
dimensionless ratio
\begin{equation}
\varepsilon = \frac{l_{\mathrm{m}}}{L_{v}} \sim \tau_{\mathrm{eff}}\left|\frac{dv}{dx}\right|.
\label{eq:22}
\end{equation}
Specifically, for \(\varepsilon \ll 1\)
the tracer experiences approximately uniform transport conditions over
the optical-memory time, and the trajectory-dependent OMT response
asymptotically reduces to the local transfer function; in contrast, for
\(\varepsilon \gtrsim 1\)
the tracer may sample appreciably varying velocities over the
optical-memory time, and the measured response may no longer be accurately represented by a single local transfer function because different
portions of the upstream trajectory contribute differently to the
accumulated optical-memory signal.

\begin{figure*}
	\includegraphics[width=1.0\textwidth]{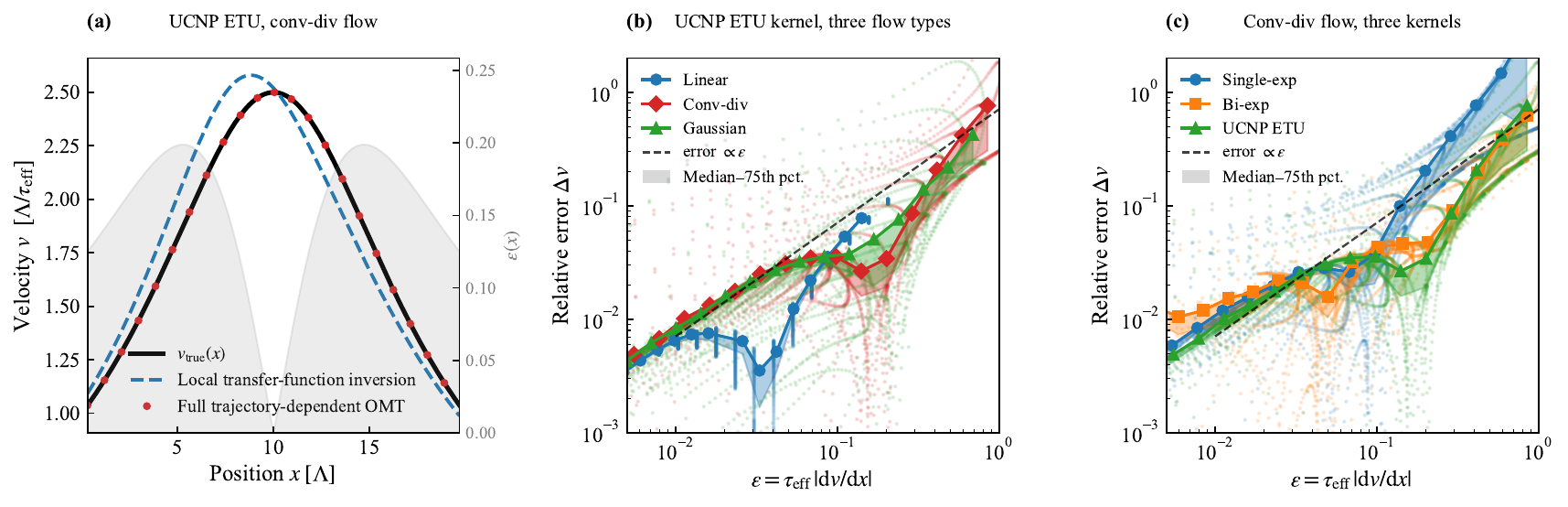}
	\caption{\textbf{Validation of the local transfer-function approximation in OMT imaging.} \textbf{(a)} Comparison between local transfer-function inversion and full trajectory-dependent inversion for a converging--diverging (conv-div) flow using the UCNP ETU kernel: true velocity \(v(x)\) (black), local transfer-function inversion (blue dashed), and full trajectory-dependent OMT inversion (red). Grey shading indicates \(\varepsilon(x) = \tau_{\mathrm{eff}}|dv/dx|\). The local inversion becomes increasingly biased in regions of large \(\varepsilon\), whereas the full trajectory-dependent inversion accurately recovers the prescribed velocity field. \textbf{(b)} Common \(\varepsilon\)-scaling of the relative velocity error \(\Delta v\) for three representative flow types (linear, conv-div, and Gaussian) for the UCNP ETU kernel. \textbf{(c)} Common \(\varepsilon\) scaling for three representative optical-memory kernels (single-exponential, bi-exponential, and UCNP ETU) using the conv--div flow. The dashed line, \(\Delta v = c\varepsilon\), illustrates the common O(\(\varepsilon\)) leading-order scaling; the proportionality coefficient depends on the memory kernel and on \(\Omega\), which varies across the data through the scanned memory times, accounting for the spread of the points. The local inversion uses kernel-specific phase look-up inversion (Supplemental Material sec. S9).}
	\label{fig:3}
\end{figure*}

To test this prediction, numerical simulations were performed to compare
the velocity reconstructed using the local transfer-function
approximation with that obtained by full trajectory-dependent OMT
inversion (Supplemental Material sec. S9), quantified by the relative
velocity error \(\Delta v = |v_{\mathrm{app}} - v|/v\), where
\(v_{\mathrm{app}}\) denotes the apparent velocity reconstructed using
the local transfer-function inversion and \(v\) is the true local
velocity. Fig.~\ref{fig:3}a shows that the full trajectory-dependent OMT inversion
accurately recovers the prescribed velocity field, whereas the local
inversion becomes increasingly biased in regions where \(\varepsilon\)
is large. Figs.~\ref{fig:3}b and 3c further show that the inversion errors
obtained from different flow types, memory times (and hence different values of \(\Omega\)),
and representative optical-memory kernels all follow the same dominant
\(\varepsilon\) scaling. This common scaling supports the prediction that the
leading-order breakdown of the local transfer-function approximation is
governed by \(\varepsilon\), whereas kernel-specific differences enter through a coefficient depending on the memory kernel, the operating parameter \(\Omega\), and the inversion procedure, determined by frequency-weighted kernel moments
(Supplemental Material sec.
S9). For the single-exponential kernel, the leading-order
phase-inversion bias can be obtained analytically and vanishes to first
order at \(\Omega = 1\) (Supplemental Material sec. S9). The analytical origin of this first-order scaling is
derived in Supplemental Material sec. S9.

The trajectory-dependent OMT response therefore provides the general
forward model for finite-memory transport, whereas the local transfer
function represents its controlled asymptotic approximation with a
quantitatively defined validity range.

\begin{figure*}
	\includegraphics[width=0.81\textwidth]{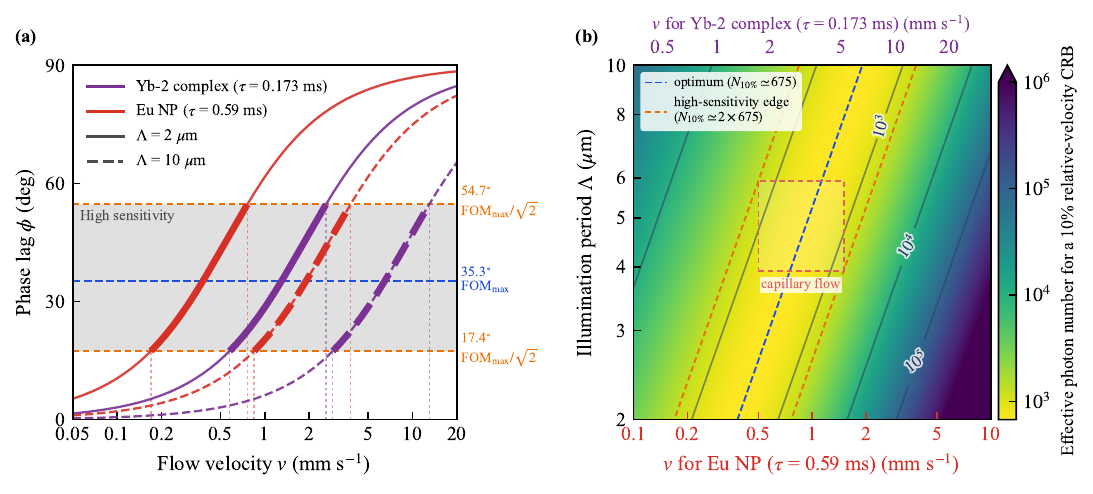}
	\caption{\textbf{Theoretical operating ranges of OMT flow imaging for two reported optical-memory tracers.} The calculations use single-exponential kernels with the reported lifetimes of Eu(III)-chelate polystyrene nanoparticles (Eu NP, 40~nm, \(\tau = 0.59\)~ms) \cite{ref13} and the Yb(III) complex Yb-2 (\(\tau = 0.173\)~ms) \cite{ref16}. \textbf{(a)} Phase lag \(\phi\) versus flow velocity \(v\) for \(\Lambda = 2~\mu\)m (solid) and \(10~\mu\)m (dashed). Thick segments and dotted projections mark the high-sensitivity range, \(\mathrm{FOM} \geq \mathrm{FOM}_{\max}/\sqrt{2}\) (grey band); the blue dashed line marks the optimum. \textbf{(b)} Effective photon number \(N_{10\%}\) [Eq.~\eqref{eq:16}] at which the CRB of Eq.~\eqref{eq:19} reaches 10\%, versus \(v\) and \(\Lambda\), read on the lower axis for Eu NP and on the upper axis for Yb-2. Blue and orange dashed lines: optimum (\(N_{10\%} \simeq 675\)) and edges of the high-sensitivity range (\(N_{10\%} \simeq 1350\)). Grey-blue lines: \(N_{10\%} = 10^{3}\), \(10^{4}\), and \(10^{5}\); values above \(10^{6}\) are saturated for display. Dashed red rectangle: the representative capillary velocity interval, 0.5--1.5~mm\,s\(^{-1}\) \cite{ref15}, and the illumination periods (about 3.9--5.9~\(\mu\)m) for which this entire interval lies within the high-sensitivity range of Eu NP. All values are calculated for idealized deterministic flow.}
	\label{fig:4}
\end{figure*}

\textbf{\textit{Experimental feasibility and operating range}}---
We now assess the experimental feasibility of OMT flow imaging. Two
practical questions are central to this assessment, namely whether the
memory times of existing tracers match velocities of practical interest,
such as those of capillary blood flow, and whether such tracers can
provide enough photons for a
quantitative readout. We address these questions with reported
optical-memory tracers, such as 40-nm Eu(III)-chelate polystyrene
nanoparticles with a
single-exponential luminescence lifetime \(\tau = 0.59\)~ms in water
\cite{ref13}, and the Yb(III) porphyrinate complex Yb-2, which is excited
in the visible, emits in the near infrared, and decays single-exponentially
with \(\tau = 0.173\)~ms in water \cite{ref16}. Both are modeled by the
single-exponential kernel of Eq.~\eqref{eq:11} with
\(\tau_{\mathrm{m}} = \tau\).

For a given tracer, the illumination period \(\Lambda = 2\pi/q\)
determines where its optical memory is most sensitive to flow. The
velocity information, measured by the FOM of Eq.~\eqref{eq:19}, is
largest at
\(v_{\mathrm{opt}} = \Omega_{\mathrm{opt}}\Lambda/(2\pi\tau_{\mathrm{m}})\),
where \(\Omega = \Omega_{\mathrm{opt}} = 1/\sqrt{2}\) [Eq.~\eqref{eq:20}]
and the phase lag is \(\phi_{\mathrm{opt}} = \tan^{-1}\Omega_{\mathrm{opt}}
\approx 35.3^{\circ}\) [Eq.~\eqref{eq:11}]. Around this optimum we define a
high-sensitivity range by \(\mathrm{FOM} \geq \mathrm{FOM}_{\max}/\sqrt{2}\),
within which the effective photon number required for a given CRB is at
most twice its minimum [Eq.~\eqref{eq:19}]. For the single-exponential
kernel this range is \(0.313 \lesssim \Omega \leq \sqrt{2}\)
(Fig.~\ref{fig:2}a), corresponding to phase lags between \(17.4^{\circ}\)
and \(54.7^{\circ}\) (grey band in Fig.~\ref{fig:4}a). For the Eu
nanoparticles at \(\Lambda = 2~\mu\)m, it spans velocities of
\(0.17\)--\(0.76\)~mm\,s\(^{-1}\). The illumination period provides a
means of adjusting this range. Because \(\Omega\) scales as \(v/\Lambda\)
[Eq.~\eqref{eq:11}], a longer period requires a proportionally higher
velocity to reach the same \(\Omega\), so increasing \(\Lambda\) to
10~\(\mu\)m shifts the range of the same tracer to
\(0.84\)--\(3.8\)~mm\,s\(^{-1}\) (Fig.~\ref{fig:4}a). The memory time of
the tracer provides a second, independent control. Because
\(\Omega \propto v\tau_{\mathrm{m}}/\Lambda\), shortening the memory time
shifts the optimum toward higher velocities, just as increasing the
illumination period does. The shorter-lived Yb-2 complex therefore
has a high-sensitivity range shifted toward higher velocities by a factor
of \(\tau_{\mathrm{Eu}}/\tau_{\mathrm{Yb}} \approx 3.4\) relative to the
Eu nanoparticles, with high-sensitivity ranges of
\(0.58\)--\(2.6\)~mm\,s\(^{-1}\) at \(\Lambda = 2~\mu\)m and
\(2.9\)--\(13\)~mm\,s\(^{-1}\) at \(\Lambda = 10~\mu\)m.

The two tracers discussed above cover about 0.2--13~mm\,s\(^{-1}\)
over illumination periods of 2--10~\(\mu\)m. Tracers with other memory
times can further extend this range. Within the Yb(III) porphyrinate family
of Yb-2 alone, chemical modification changes the memory time by a factor
of three \cite{ref16}, and other reported lanthanide tracers offer memory
times from about 0.06 to 1.5~ms (Supplemental Material Table~S2).
Selecting among such tracers extends the measurable velocities from below
0.1 to about 40~mm\,s\(^{-1}\) over the same range of illumination
periods (Supplemental Material Table~S2), which offers the flexibility to meet the requirements of
different flow applications.

A favorable sensitivity range is useful only if the corresponding
precision can be reached with an attainable number of photons. To
quantify this photon cost, we translate the CRB of Eq.~\eqref{eq:19} into
the effective photon number [Eq.~\eqref{eq:16}] at which the relative
velocity standard deviation reaches 10\%,
\(N_{10\%} = 100/\mathrm{FOM}^{2}\), and map it in Fig.~\ref{fig:4}b as a
function of \(v\) and \(\Lambda\). The photon cost is lowest along the
optimum \(\Omega = \Omega_{\mathrm{opt}}\), where Eq.~\eqref{eq:20} gives
\(N_{10\%} \simeq 675\). At the edges of the high-sensitivity range,
where by definition the FOM has fallen to \(\mathrm{FOM}_{\max}/\sqrt{2}\),
it doubles to \(N_{10\%} \simeq 1350\). Beyond them it rises steeply,
most rapidly on the high-velocity side, where both the modulation
amplitude and the slope of the phase response decrease. Because the
photon cost, like the phase response, depends on velocity, memory time,
and illumination period only through \(\Omega\), a single map serves both
tracers once the velocity axis is rescaled by the memory time, and the two
velocity axes of Fig.~\ref{fig:4}b show the corresponding velocity ranges
for the Eu and Yb-2 tracers.

Whether an existing tracer can supply these photon numbers under
representative imaging conditions is the remaining question. We estimated
the photon budget of a representative wide-field measurement from the
reported chelate loading, quantum yield, and lifetime of the Eu
nanoparticles \cite{ref13} (Supplemental Material sec. S11). Within a
1-ms acquisition window, tracer concentrations of about 25--50~nM supply
the effective photon numbers of Fig.~\ref{fig:4}b, from the optimum to
the edges of the high-sensitivity range. This order-of-magnitude estimate
indicates that the photon numbers required for the precision targets of
Fig.~\ref{fig:4}b are within reach of existing tracers.

These operating ranges are directly relevant to microvascular blood
flow. In rat cerebral capillaries, Ivanov et al. measured a mean
red-cell velocity of about 0.8~mm\,s\(^{-1}\), with 63\% of the examined
capillaries showing mean velocities between 0.5 and 1.5~mm\,s\(^{-1}\)
\cite{ref15}. For tracers with optical-memory properties similar to those
of the Eu nanoparticles, this entire interval falls within the
high-sensitivity range defined above for illumination periods between 3.9 and
5.9~\(\mu\)m (dashed rectangle in Fig.~\ref{fig:4}b).

\textbf{\textit{Discussion}}--- The present theory
establishes OMT imaging as a measurement framework in which the
fundamental observable is the transport history rather than the
instantaneous local flow state. Unlike conventional motion blur or
trailing effects, whose nonlocality originates from the acquisition
process itself, the transport-history dependence derived here arises
from the physical memory of the tracer. The measured optical signal
therefore remains spatially localized at the detection coordinate while
being determined by the upstream transport history sampled during the
optical-memory time. OMT imaging therefore measures a complex
transport-history response jointly determined by the flow field and the
optical-memory kernel, rather than the instantaneous local velocity
itself.

Under locally uniform transport, the trajectory-dependent OMT response
reduces systematically to the local transfer-function description,
enabling quantitative velocity inversion and information-theoretic analysis. The
dimensionless parameter
\(\varepsilon = \tau_{\mathrm{eff}}\left|dv/dx\right|\),
comparing the transport distance sampled during the optical-memory time
with the characteristic streamline-variation length, separates two
distinct operating regimes. For \(\varepsilon \ll 1\), the local
transfer-function approximation accurately describes the measured
response, permitting efficient local velocity inversion. When
\(\varepsilon \gtrsim 1\), however, the accumulated upstream transport
history may no longer be accurately represented by a single local transfer
function. Quantitative flow mapping nevertheless remains feasible
through direct inversion of the trajectory-dependent forward model using
a calibrated optical-memory kernel together with an appropriate
parametric flow description.

The linear response formulation applies to both individual tracers and
spatially uniform tracer ensembles. For ensemble measurements, the
present analysis assumes that the tracer concentration is sufficiently
uniform that the measured spatial response is governed primarily by
transport through the optical-memory kernel. More generally, spatial
concentration heterogeneity can be incorporated into the forward model
through an appropriate description of the concentration field, enabling
quantitative transport analysis beyond the uniform-concentration
approximation.

For complex nonlinear tracers such as UCNPs, quantitative OMT imaging
does not require explicit identification of all microscopic rate
constants. In practice, the operating-point-dependent complex transfer
function can be calibrated directly as
\(H_{\mathrm{cal}}\left( qv;I_{0} \right)\) under known uniform flow and
subsequently used for velocity inversion at the same excitation
operating point. The microscopic ETU model provides physical
interpretation and guides kernel engineering, whereas the calibrated
transfer function serves as the practical forward model for imaging.

In OMT, the operating range is not fixed by the imaging
architecture alone but is determined jointly by the memory time of the
tracer and the illumination period. The tracer memory time is therefore a
key parameter for selecting the velocity range, and measuring faster
flows, as in larger microvessels, calls for shorter memory times. At \(\Lambda = 5~\mu\)m, for example,
\(\tau \approx 56~\mu\)s places \(v = 10\)~mm\,s\(^{-1}\) at the optimum,
and a memory time of this order has already been reported for a
water-soluble Yb(III) porphyrinate \cite{ref16}. Luminescent materials
with such memory times are more widely available than millisecond
emitters, which broadens the choice of suitable tracers. Kernel
engineering offers a complementary route. By changing the shape of the
transfer function rather than its time scale, it may extend the useful
velocity range beyond that of a single-exponential response.

Beyond the memory time, the excitation and emission wavelengths
also influence the suitability of a tracer for a given imaging
application. The Eu tracer considered here
requires UV excitation, which limits its use, in particular in tissue,
whereas the Yb-2 complex is excited in the visible and emits in the near
infrared \cite{ref16}. Because OMT requires only a finite optical memory
matched to the illumination period and flow velocity, together with
sufficient brightness, these properties can be chosen separately across a
wide range of luminescent materials. UCNPs, for instance, are excited in
the near infrared and have
memory times that can be tuned through their composition and excitation
conditions \cite{ref3,ref6}. Tracers with different memory times and
spectrally separable emissions could, in principle, also be combined in
one measurement and detected in separate spectral channels, covering
adjacent velocity ranges simultaneously, provided that suitable excitation
conditions are available.

The present framework also places our previously proposed PP-SIV \cite{ref3} on a rigorous
theoretical foundation. PP-SIV implicitly operated in the local
constant-velocity limit of the present trajectory-dependent formulation,
corresponding to the asymptotic regime \(\varepsilon \ll 1\). Within the
OMT framework, the previously proposed look-up inversion and full
parametric fitting emerge naturally as the local transfer-function
approximation and the full trajectory-dependent inversion, respectively.

Finite optical memory has long been exploited, explicitly or implicitly,
in fluorescence spectroscopy and imaging utilizing the memory of the
excited-state ($\sim$ns) \cite{ref2} or the memory of the
long-lived dark-states (monitored via fluorescence blinking)
\cite{ref9,ref10,ref11}. The present work establishes a unified transport-response
framework that extends these finite-memory concepts to quantitative
transport imaging.

Although the present derivation is motivated by nonlinear photophysical
rate equations, the resulting framework extends well beyond luminescent
tracers. The resulting transport framework applies generally to any
finite-memory observable advected through a spatially structured
excitation field, irrespective of whether the memory originates from
photophysical relaxation, chemical kinetics, or other finite-memory
processes.

The present analysis focuses on deterministic transport
and does not explicitly account for stochastic tracer displacements,
which may affect the measured response under experimental conditions.
The trajectory-integral formulation, however, is not restricted to deterministic
advection. Replacing deterministic trajectories by the corresponding
stochastic propagators naturally extends the theory to diffusive,
dispersive, and more general stochastic transport processes. This
stochastic extension has been developed to describe scalar and tensor diffusion
imaging in a separate preprint~\cite{ref14}. For nanoparticle
tracers of tens of nanometers, the diffusive displacement during the
memory time, \(\sqrt{2D\tau_{\mathrm{m}}}\), with \(D\) the Stokes--Einstein diffusion
coefficient, is about 0.1--0.2~\(\mu\)m, small compared with the
illumination periods considered here, so that diffusion has only a minor
effect on the measured response.

The present work therefore establishes OMT as a general theoretical
framework for finite-memory transport imaging, in which
transport-history encoding, quantitative inversion, information-optimal
measurement, and kernel engineering are unified within the same
transport-response formalism. Structured illumination provides one
particularly convenient implementation, but the underlying theory is
independent of the specific excitation geometry. More broadly, the OMT
framework provides a general strategy for designing finite-memory probes
and imaging modalities that quantify transport through the controlled
interplay between tracer memory and structured excitation. OMT therefore
establishes a general measurement framework in which finite tracer
memory becomes a quantitative observable for transport imaging.

\begin{acknowledgments}
\textbf{\textit{Acknowledgments}}--- The authors used OpenAI ChatGPT (GPT-5.6 Sol)
and Anthropic Claude (accessed through Claude Code)
during manuscript preparation as auxiliary tools for consistency
checking, verification of mathematical derivations and numerical
results, and refinement of scientific exposition. AI-assisted
suggestions were critically evaluated and, where relevant,
independently checked against the analytical derivations, numerical
calculations, and source code by the authors. The authors take full
responsibility for all scientific content and conclusions. H.L. and
J.W. acknowledge financial support from the ÅForsk Foundation
(23-322), the Carl Trygger Foundation (23-2635), and the Swedish
Research Council (VR 2021-04556, 2025-05609).
\end{acknowledgments}

\textbf{\textit{Data availability}}--- The data and
numerical results supporting the findings of this study are available
from the corresponding author upon reasonable request.

\bibliography{references}

\end{document}